# Virtual Histological Staining of Label-Free Total Absorption Photoacoustic Remote Sensing (TA-PARS)

**Marian Boktor[1], Benjamin R. Ecclestone[1,2], Vlad Pekar[1,2], Deepak Dinakaran[3], John R. Mackey[3], Paul Fieguth[4], Parsin Haji Reza[1,*]**

[1]PhotoMedicine Labs, Systems Design Engineering, University of Waterloo, Waterloo, Ontario, N2L 3G1, Canada
[2]illumiSonics, Inc., 22 King Street South, Suite 300, Waterloo, Ontario, N2J 1N8, Canada
[3]Cross Cancer Institute, Department of Oncology, University of Alberta, 116 St & 85 Ave, Edmonton, Alberta, T6G 2V1, Canada
[4]Systems Design Engineering, University of Waterloo, Waterloo, Ontario, N2L 3G1, Canada
* phajireza@uwaterloo.ca

## Abstract

Histopathological visualizations are a pillar of modern medicine and biological research. Surgical oncology relies exclusively on post-operative histology to determine definitive surgical success and guide adjuvant treatments. The current histology workflow is based on bright-field microscopic assessment of histochemical stained tissues and has some major limitations. For example, the preparation of stained specimens for brightfield assessment requires lengthy sample processing, delaying interventions for days or even weeks. Therefore, there is a pressing need for improved histopathology methods. In this paper, we present a deep-learning-based approach for virtual label-free histochemical staining of total-absorption photoacoustic remote sensing (TA-PARS) images of unstained tissue. TA-PARS provides an array of directly measured label-free contrasts such as scattering and total absorption (radiative and non-radiative), ideal for developing H&E colorizations without the need to infer arbitrary tissue structures. We use a Pix2Pix generative adversarial network (GAN) to develop visualizations analogous to H&E staining from label-free TA-PARS images. Thin sections of human skin tissue were first virtually stained with the TA-PARS, then were chemically stained with H&E producing a one-to-one comparison between the virtual and chemical staining. The one-to-one matched virtually- and chemically- stained images exhibit high concordance validating the digital colorization of the TA-PARS images against the gold standard H&E. TA-PARS images were reviewed by four dermatologic pathologists who confirmed they are of diagnostic quality, and that resolution, contrast, and color permitted interpretation as if they were H&E. The presented approach paves the way for the development of TA-PARS slide-free histological imaging, which promises to dramatically reduce the time from specimen resection to histological imaging.

## Introduction

Histopathology is the fundamental microscopic examination tool in many domains including cancer diagnosis and prognosis, surgical oncology, and drug discovery and development[1]. Tissue staining for microscopic imaging enabled histopathology by helping to distinguish the composition of tissue samples. The gold standard of histological imaging is brightfield microscopy of stained formalin fixed paraffin embedded (FFPE) thin tissue preparations, where the most common staining set is Hematoxylin and Eosin (H&E). Hematoxylin stains cell nuclei with deep blue-purple, while eosin stains cytoplasm and extracellular matrix with pink shades[2].

Developing these stained FFPE specimens for brightfield microscopic assessment relies on a laborious process of fixation, embedding, sectioning, and staining[3], a procedure which takes several days to complete[4]. This lengthy process has motivated the adoption of frozen section (FS) histology, a technique which is commonly used for rapid intraoperative assessment. FS enables turnaround times of about 20 minutes[5]. However, sampling of specimen for FS is limited and involves difficult technical processes. Rapid freezing introduces artifacts deteriorating the cellular morphology, which in turn affects the pathologist diagnosis of tissue samples by pathologists[5]. Due to these limitations, FFPE histochemical procedures remain the irreplaceable gold standard method.

Histochemical staining not only delays sample assessment, but the sample preparation steps permanently alter the tissue composition with multiple chemical substances, meaning tissues may only be stained once per section[2]. An independent slide must be prepared for each desired staining contrast. This intensive FFPE tissue preparation may delay cancer diagnoses and margin assessments for over a week, which in turn delays adjunct treatment and supplementary procedures. This motivates a clinically acceptable technique to provide accurate, fast, and reproduceable histology-*like* images label free.

Since pathologists are primarily trained to make diagnoses on histochemical stained tissue samples, one promising approach is to develop algorithms that generate virtual H&E-*like* visualizations of unlabeled tissue, simulating the chemical staining effects. This will ease the adoption of label-free tools. In addition, virtual staining preserves tissues, permitting downstream processing of the same tissue sections. In a clinical setting this would improve the diagnostic utility of small tissue specimens. Moreover, virtual stains, by eliminating pre-analytic variability, should reduce inter- and intra-laboratory staining variability caused by slight process variations, thus addressing a major challenge in clinical pathology. Finally, directly replicating current staining enables one-to-one validation against true H&E, a necessary step in adoption.

Several optical microscopes have been developed to provide H&E-like imaging capabilities, such as quantitative phase imaging[6], reflectance confocal microscopy (RCM)[7], microscopy with ultraviolet surface excitation (MUSE)[8,9], optical coherence tomography (OCT)[10–12], autofluorescence microscopy[13], and photoacoustic microscopy (PAM)[14–17]. These methods are usually coupled with deep learning-based virtual staining models, particularly generative adversarial networks (GANs)[18]. Microscopic images are captured and then are used to train a network to perform virtual H&E staining. Hence, the quality of the resulting virtual H&E staining will be directly determined by the characteristics of the optical microscope used to capture the images. In the ideal case, the microscopy technique would (1) operate in reflection mode, enabling imaging of thick tissue specimens, and (2) would provide analogous contrast to current histochemical staining, improving artificial colorization quality by avoiding inferences of essential diagnostic features. Other optical microscopes that have done distinguished progress in H&E-*like* imaging include light-sheet microscopy (LSM)[19] and stimulated Raman spectroscopy (SRS)[20,21].

Of the listed techniques, MUSE and LSM have demonstrated acceptable agreement with H&E-stained histology, mimicking the standard H&E staining specificity. However, both MUSE and LSM rely on fluorescence stains to provide H&E contrast. These dyes can potentially be mutagenic, carcinogenic, or toxic, hindering *in-situ* applicability of these fluorescence-based techniques[9]. Moreover, LSM requires relatively clear samples for 3D volumetric imaging of fresh tissue, which necessitates additional equipment and long preprocessing times making the technique impractical in an intraoperative setting[19]. Accordingly, label-free techniques are expedient.

One popular label-free modality, SRS, leverages the vibrational resonance of molecular bonds to provide contrast[22]. SRS has demonstrated promising results when compared to H&E stains in both thin and thick tissue samples. However, SRS usually operates in transmission mode[21–23]. Thick tissues are typically compressed to transmissible thicknesses to acquire images[20], which severely alters tissue morphology. This not only affects the pathologist assessment, but also renders SRS impractical for *in-situ* imaging.

Conversely, OCT usually operates in reflection mode, and is well suited to imaging thick tissue specimens. OCT has shown potential in label-free tissue block imaging[11,24], surgical margin assessment[25,26], and biopsy examination[27,28]. However, OCT uses optical scattering contrast which highlights predominately morphological information. This contrast lacks the sufficient specificity to distinguish biomolecules like cell nuclei and cytoplasm, which is essential for histological analysis. Subsequently, artificial colorization methods must infer the structures of these essential diagnostic features, which greatly reduces diagnostic confidence. This motivates the adoption of a label-free microscopy modality with higher chromophore specificity, which may improve colorization confidence.

Recently, Autofluorescence microscopy has emerged as a popular contender for virtual histology. Autofluorescence provides label-free visualizations of biomolecules like elastin, collagen, amino acids, NADPH and other cellular organelles, at excitation wavelengths ranging from UV to ~500 nm[29]. Previously, Rivenson, Y. *et al.*[13] had success in developing virtual H&E staining from autofluorescence images. In this work, a deep neural network was trained on pairs of autofluorescence and H&E-stained images with the objective of colorization. This method was successfully applied to autofluorescence images, bypassing the H&E staining process. However, the broad realization of this technique in bulk tissues was limited by the autofluorescence contrast. Though autofluorescence highlights a number of biomolecules resembling eosin staining, the DNA and nuclei[30], do not exhibit any measurable autofluorescence[31]. Hence, while many diagnostic elements are present, the colorization network must estimate the nuclear features since they cannot be measured. Inferring nuclear structures (a key determinant of cancer diagnosis) presents challenges to clinical adoption and regulatory approval.

One modality, PAM has arisen as a powerful label-free microscopy modality, offering selective biomolecule contrast of a wide array of chromophores. PAM enables precise discrimination by using specific excitation wavelengths to target the optical absorption characteristics of individual chromophores. Previously, PAM systems have been shown to recover both nuclear visualizations analogous to hematoxylin staining and connective tissue imaging analogous to eosin staining[15]. However, traditional PAM systems exhibit one major drawback. PAM is a hybrid optoacoustic imaging modality, where optical absorption induced photoacoustic pressures are captured as ultrasound signals. Measuring these acoustic pressure waves requires an acoustically coupled ultrasound transducer. The need for effective acoustic coupling, means the transducer must be in contact with the specimen, or submerged in a coupling medium such as a water tank[14–16]. This arrangement hinders the suitability of PAM for several clinical applications.

A newly developed modality, photoacoustic remote sensing (PARS)[32–34], overcomes the drawbacks of PAM. In PARS, the acoustically coupled ultrasound transducer is replaced with a secondary co-focused detection laser, offering robust all-optical, non-contact, label-free photoacoustic imaging[35]. Over the past few years, PARS has emerged as a powerful contender in the label-free histological imaging space[32–34,36]. PARS may emulate current histochemical staining, by targeting the UV absorption of DNA (analogous to hematoxylin staining)[32,33], and by targeting the 420 nm absorption of cytochrome (analogous to hematoxylin staining)[32]. Recent works have shown that PARS may offer high contrast and high spatial resolution in freshly resected tissue, formalin fixed tissue, and FFPE tissue sections[33,34]. In resected tissues, PARS has even been shown to provide 3D imaging of subsurface nuclei [33].

In this paper, we employ a second-generation PARS architecture, total-absorption photoacoustic remote sensing (TA-PARS)[34], to provide label-free histology-*like* images. As outlined by Ecclestone *et al.* the TA-PARS architecture incorporates a number of advances providing improved sensitivity and contrast when imaging in tissues specimen[34]. Applied to resected tissues, TA-PARS directly measures analogous contrast to H&E staining, capturing nuclei and cytoplasm independently. Since the proposed framework directly measures H&E information, it enables confident colorization modeling. In addition, this provides the opportunity for direct comparison to ground truth H&E staining while providing outputs acceptable for clinical applications.

The TA-PARS operates by introducing a pulse of targeted excitation light to a chromophore, then capturing the relaxation processes as the chromophore sheds the absorbed optical energy. There are two main pathways for this relaxation: radiative, and non-radiative. During non-radiative relaxation, absorbed optical energy is dissipated as heat. This generates local thermoelastic expansion, and photoacoustic pressures when the excitation process is fast enough[35]. The non-radiative relaxation induces modulations in the local optical properties, which are captured as back-reflected intensity fluctuations in a co-focused TA-PARS detection laser. When using a 266 nm excitation, the non-radiative absorption reveals predominantly nuclear structures, typically visualized by the "H" stain. Conversely, during radiative relaxation, absorbed energy is emitted as photons. The radiative relaxation is captured as these absorption-induced optical emissions, the same mechanism as fluorescence imaging. When using a 266 nm excitation, the radiative absorption contrast typically depicts extranuclear features, commonly revealed by the "E" stain. In addition to the optical absorption fractions, the local scattering is also captured using the TA-PARS detection laser. The optical scattering reveals the morphological structures in thin tissue sections[36].

In total, TA-PARS microscopy simultaneously provides label-free non-contact imaging of radiative absorption, non-radiative absorption, and optical scattering contrasts. To the extent of our knowledge, the TA-PARS is the only modality that can provide all three contrast mechanisms in a single capture. This rich array of input data provides simultaneous recovery of structures including cell nuclei, fibrin, connective tissues, adipocytes, and neurons[34] in a single excitation event. The diagnostic confidence of deep learning-based virtual histological staining can be improved using this more informative dataset for network training and colorization.

Here, we use a general-purpose image-to-image translation model, namely Pix2Pix[37], to colorize histological images captured by the TA-PARS system. The Pix2Pix model is designed based on generative adversarial networks (GANs)[18], a type of generative model that creates new datasets based on a given input (training) data. The architecture is comprised of a generator subnetwork for creating new feasible data, and a discriminator subnetwork that tries to distinguish between the newly generated (synthetic) data and the ground truth (real) data. The two subnetworks are trained simultaneously in an adversarial process, where the generator model aims to maximize the error rate of the discriminator model, while the discriminator model tries to minimize the classification error. For the training task, the network was fed multi-contrast images of the TA-PARS channels co-registered with H&E images that were captured using traditional slide imaging procedures.

Through the array of contrasts provided by TA-PARS, we directly capture the H&E information. Directly measuring the desired H&E contrast in one dataset permits a strong deep learning model, avoiding assumptions or unsubstantiated inferences about the presence of features such as cell nuclei. The proposed work provides H&E-*like* TA-PARS of tissue slides, leading to images comparable to the gold standard. TA-PARS virtual histology images were reviewed by four dermatologic pathologists who confirmed they are of diagnostic quality, and that resolution, contrast, and color permitted interpretation as if they were H&E. This is an essential step toward developing an alternative histopathological workflow featuring slide-free virtual staining of fresh tissue specimens. Ultimately, achieving histology-like *in-situ* imaging would permit near real-time intra-operative margin assessment.

## Materials and Methods

### Dataset Preparation

The dataset used in this study was collected from thin sections of formalin fixed paraffin embedded (FFPE) human skin tissues. Anonymous tissues samples were provided by clinical collaborators at the Cross-Cancer Institute (Edmonton, Alberta, Canada). These samples were collected under protocols approved by the Research Ethics Board of Alberta (Protocol ID: HREBA.CC-18-0277) and the University of Waterloo Health Research Ethics Committee (Photoacoustic Remote Sensing (PARS) Microscopy of Surgical Resection, Needle Biopsy, and Pathology Specimens; Protocol ID: 40275). Patient consent was waived by the ethics committee as the samples are archival tissues no longer required for patient diagnostics, and no information was provided to the researchers about the patient identity.

The dataset of co-registered TA-PARS and H&E pairs required for the virtual staining procedure was acquired as follows. Unstained thin tissue sections were imaged with the TA-PARS system. Once imaged, specimens were stained and scanned with a brightfield microscope to generate matched TA-PARS and H&E image pairs of the exact same samples. The dataset preparation procedure is illustrated in Fig. 1.

The unstained tissue sections were prepared as follows. First, bulk resected tissues were fixed in a formalin fixative solution (within 20 minutes of resection) for up to 48 hours. Tissues were then dehydrated and prepared for paraffin wax penetration. During this process, tissues were cleared of residual fats. Next, tissues were embedded into the paraffin substrate creating FFPE tissue blocks. Thin sections of ~5 µm thickness were sliced from the FFPE tissue block surface. Thin sections were fixed directly onto glass slides and excess paraffin was removed by baking the slide at 60 °C for 60 minutes. At the end of this process, the thin sections were ready for imaging with the TA-PARS system. All imaging was done at the PhotoMedicine Labs at the University of Waterloo. Once the TA-PARS dataset was collected, the thin tissue slides were stained with H&E, covered with a mounting medium and a coverslip, and then imaged using a transmission mode brightfield microscope, providing the corresponding H&E dataset.

### TA-PARS Imaging

A detailed description of the TA-PARS system architecture and image formation process can be found in[34]. Briefly, the experimental system is illustrated in Fig. 2. The excitation source was a 266 nm diode laser (Wedge XF 266, RPMC). The detection was a 405 nm OBIS-LS laser (OBIS LS 405, Coherent). The excitation and detection sources were combined via dichroic mirror, then co-focused onto the sample with a 0.42 NA UV objective lens. Detection light returning from the sample (containing the optical scattering and non-radiative relaxation) was fiber-coupled into the circulator, which then redirected the reflected light to a photodiode where the nanosecond scale photoacoustic intensity modulations were captured. Concurrently, radiative relaxation was chromatically isolated and then captured using a photodiode.

The TA-PARS image formation process can be described as follows. Mechanical stages were used to scan the sample while the objective lens remained fixed. The velocity of the stages was adjusted to obtain a pixel size of 250 nm. The 50 kHz pulsed excitation source was used to excite the tissue, and the induced signals due to optical scattering, radiative, and non-radiative relaxation were collected at each pulse. Each event was placed in the channel that corresponds to its signal. Such signals were compressed by extracting a single feature that becomes the pixel value. An image was then reconstructed for each TA-PARS channel through fitting the computed pixel values to a Cartesian grid based on their corresponding position signals from the stages. The images were then processed according to the task of interest. In the next subsection, processing of images for the virtual staining task is described.

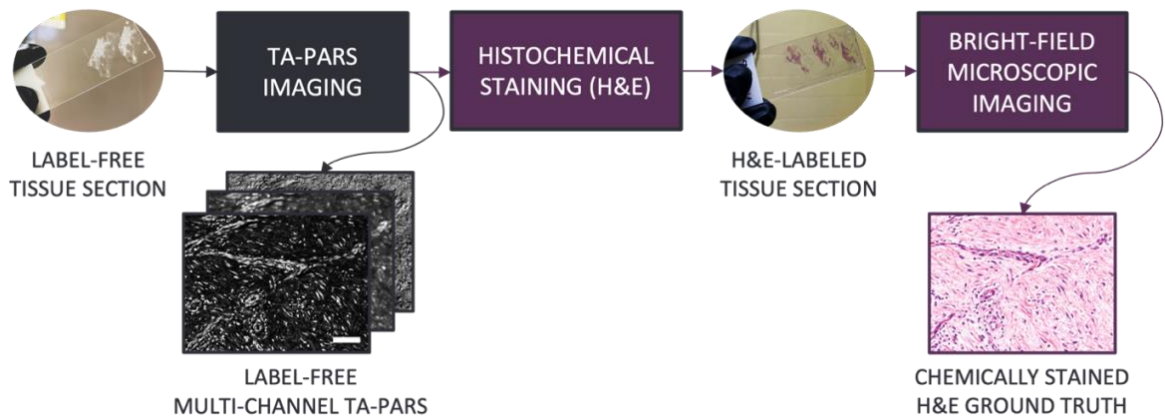

**Figure 1.** TA-PARS and H&E dataset preparation process of human skin tissue slide images. Scale bar: 50 µm.

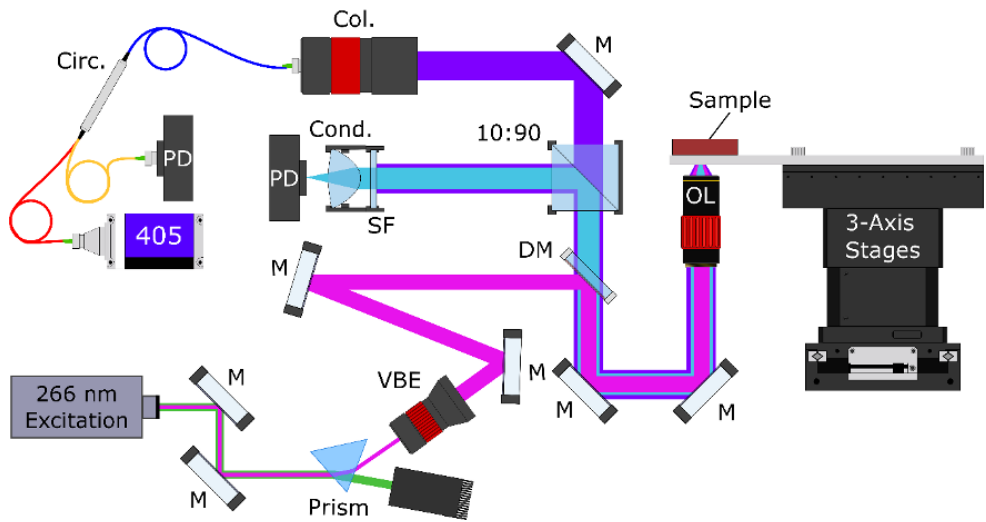

**Figure 2.** Simplified TA-PARS system architecture and system setup. Component labels are as follows: dichroic mirror (DM), variable beam expander (VBE), collimator (Col.), circulator (Circ.), spectral filter (SF), condenser lens (Cond.), photodiode (PD), 10:90 splitter (90:10), mirror (M) and objective lens (OL).

## Dataset Preparation for Training

In this work, a Pix2Pix GAN model[37] was adopted. The model requires pixel-to-pixel correspondences between the input data (TA-PARS) and the ground truth (H&E), where the objective is to learn a statistical transformation between the two data domains. Since TA-PARS and H&E images were acquired using different techniques, images of the same sample were not co-registered. Therefore pre-processing, including field of view matching and registration, was required to produce approximately co-registered pairs of TA-PARS and H&E images. The virtual staining framework is depicted in Fig. 3.

The registration process can be described as follows. First, the corresponding field-of-views (FOVs) of the two image domains were extracted from the whole-slide images and coarsely matched in terms of pixel size in preparation for one-to-one registration process. For registration, the control point registration tool was deployed from the MATLAB Image Processing Toolbox. The TA-PARS images were chosen as reference images, and H&E as the images to be registered. The control points were manually selected and then fine-tuned in a second pass to minimize global and local distortions. Afterward, the algorithm fits a non-rigid geometric transformation[38] between the two images, which was applied to the H&E images to obtain co-registered TA-PARS and H&E images. Since all TA-PARS channels were intrinsically registered, the whole registration process was applied using the non-radiative channel only.

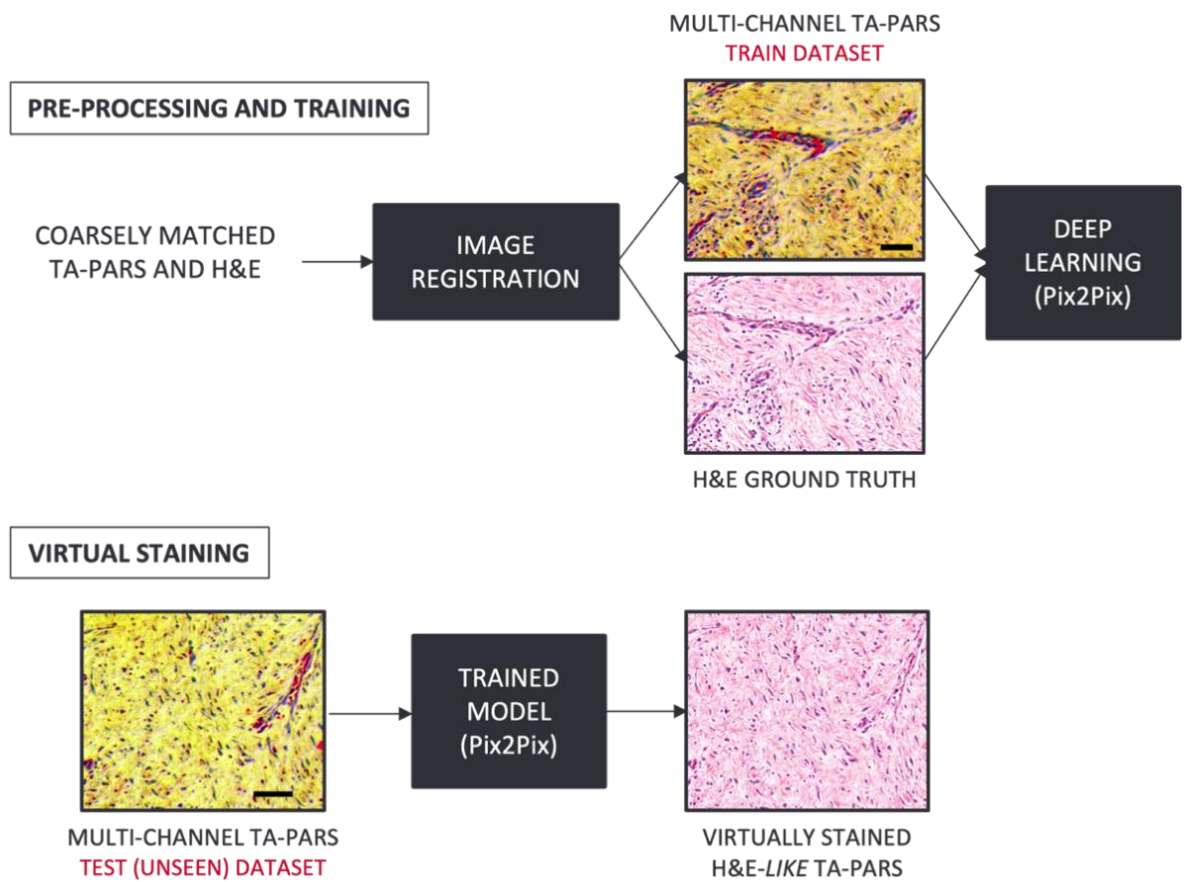

**Figure 3.** Virtual Staining framework. Pairs of TA-PARS and H&E images are co-registered to create the training dataset, which is then used to train the Pix2Pix[37] model. Unseen TA-PARS images are then fed into the trained model for virtual staining, producing H&E-like visualizations. Scale bar: 50 μm.

Once the registration process was done, the dataset was ready for the network training for colorization. In this study, all three TA-PARS channels were used for training the model. Figure 4 illustrates the inputs and the output of the network. Non-radiative and radiative absorption and optical scattering channels are shown in Fig. 4(a-c), respectively. The multi-channel input is shown in Fig. 4(d) as the concatenation of the three channels. A colorization example is depicted in Fig. 4(e) and the corresponding H&E ground truth is shown in Fig. 4(f). It is clear that the non-radiative absorption channel features strong nuclear contrast representing the hematoxylin contrast, while the radiative absorption contrast exhibits the eosin contrast showing the tissue morphological information. Normalized TA-PARS images undergo contrast stretching by saturating the top 1% and the bottom 1% of all the pixel values and were then color-reversed to match the colormap and the histogram distribution of the grayscale ground truth images. Several datasets were prepared in a likewise manner for concept validation.

This framework is meant to replace the histological staining process, where the Pix2Pix model was trained using the multi-channel TA-PARS images as inputs and corresponding H&E images as labels, and the network was trained to learn the statistical transformation function between the two domains. Once the model is trained, the virtual staining process takes a few seconds to produce a virtually stained version of an image of 4000 x 6500 pixels (~1 mm$^2$).

## Virtual Staining Model Architecture and Implementation Details

A Pix2Pix image-to-image translation model[37] was applied to transform the label-free TA-PARS images to virtually stained images that match their corresponding H&E histochemical stained samples. The model is based on a GAN architecture, or more specifically conditional GAN (cGAN)[39], where the generative model is further conditioned on the input data.

The model training algorithm in Pix2Pix concurrently trains two separate neural networks, the generator, and the discriminator, as illustrated in Fig. 5. The training algorithm learns to minimize a loss function between the network output and the ground truth. The discriminator network is trained to effectively differentiate between real and synthetically generated images, while the generator network is trained at the same time to produce images that better resemble the ground truth data, which makes it progressively harder for the discriminator to differentiate correctly. As a result, the model is capable of producing virtually stained TA-PARS images of a comparable diagnostic quality with the histologically stained images.

Training for the standard Pix2Pix model was carried out using 3-channel RGB color images. In the proposed approach, the three channels were replaced with the non-radiative absorption contrast, radiative absorption contrast, and the optical scattering, respectively. In this way, the model can simultaneously learn complementary information from the three available information sources.

Overlapping patches of 256x256 pixels were extracted from TA-PARS and H&E images to generate the training and validation sets. Approximately, 15000 training patches and 6000 validation patches were used in the experiments. The maximum number of epochs was set to 500 with an early stopping criterion to terminate the training when the generator loss stops improving. The trained model was then applied to the test images which were also subdivided into overlapping patches of 256x256 pixels. An overlap of ~50% was usually sufficient to avoid visible artifacts at the borders of adjacent patches in the final stitched image. The colorization algorithm was implemented in Python version 3.8.10 and model training was implemented using PyTorch version 1.9.1 with support of CUDA version 11.

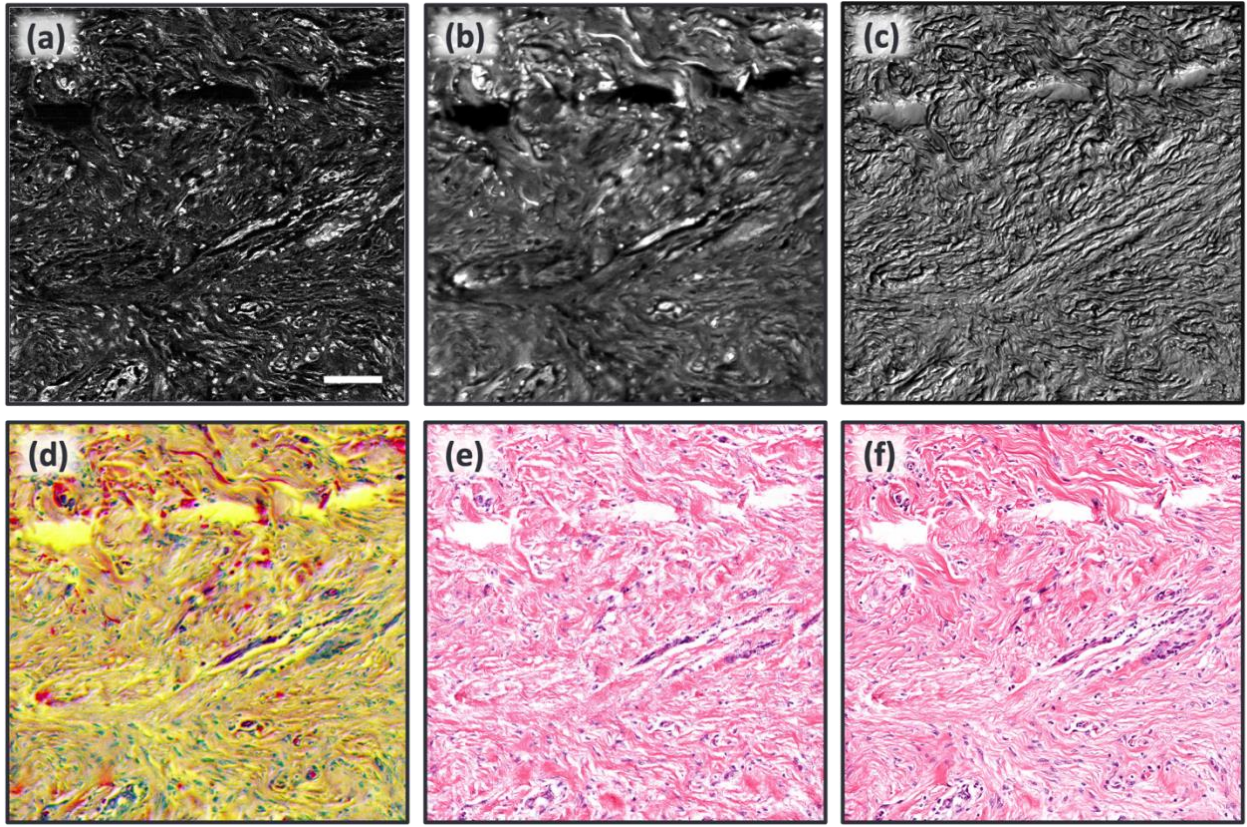

**Figure 4.** The colorization network input and output images of human skin tissue slide. (a) TA-PARS non-radiative absorption contrast image captured using a TA-PARS microscope. (b) TA-PARS radiative absorption contrast of the same tissue using the same microscope. (c) TA-PARS optical scattering contrast. (d) TA-PARS image of all three channels (network input to be colorized). (e) Colorized TA-PARS image using Pix2Pix model. (f) Histologically stained tissue of the same region taken on a bright field microscope, the ground truth for the virtual staining framework. Scale bar: 100 μm.

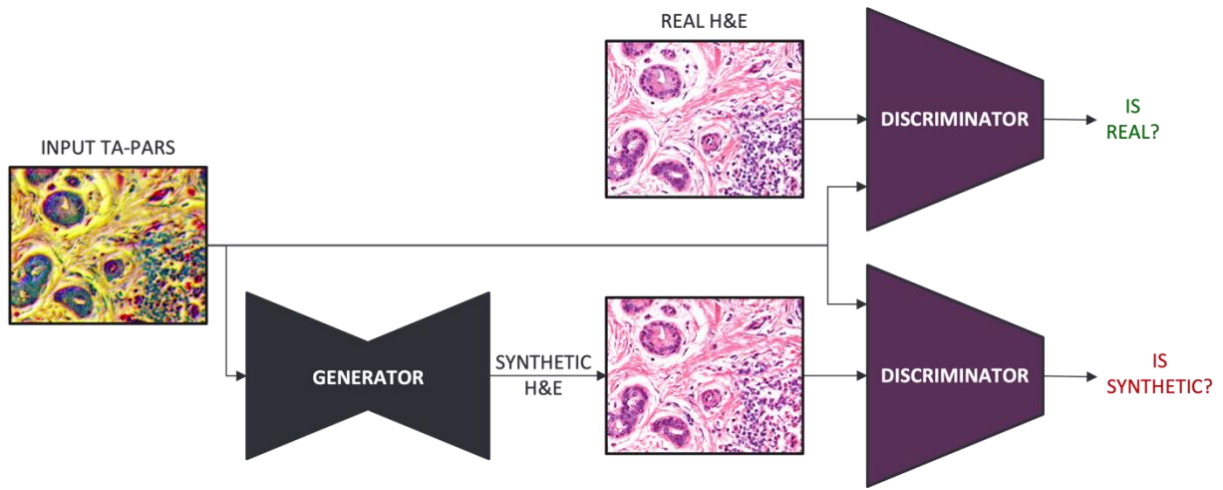

**Figure 5.** An illustration of the Pix2Pix training algorithm. Pix2Pix is comprised of two subnetworks: the discriminator and the generator. The discriminator subnetwork is trained to distinguish between real H&E and synthetically generated H&E. Simultaneously, the generator subnetwork is trained to confuse the discriminator by producing images that looks like real H&E.

## Clinical Validation of Virtual H&E Images of Skin

Nine representative images of TA-PARS generated, and AI-colorized human skin were distributed to four board certified dermatopathologists for evaluation. None of the dermatopathologists had conflicts of interests or financial relationships relating to this technology. Each pathologist confirmed that all the images are of sufficient quality to allow diagnosis, and that contrast, resolution, and colorization quality are sufficiently H&E-*like* to permit clinical diagnoses.

## **Results and Discussion**

We applied our machine learning framework to unstained human skin tissue sections imaged with TA-PARS. These results were compared against bright-field microscopy images, which were acquired after staining the exact same tissues with H&E. Hence, these images provide a one-to-one nuclei-to-nuclei match between the TA-PARS, and corresponding H&E images. Given limited datasets, our initial step was to perform a proof of concept and bracket the model's performance by overfitting. When overfitting, the same data was used for training and testing the model. This was performed using a single image, split into 5000 patches of 256 x 256 pixels each. The colorization results of this experiment are visually almost indistinguishable from the ground truth images (as shown in Fig. 6), setting an upper bound to the model performance. When generalizing the model, comparable results can be realistically achieved with data not included in the training dataset (unseen data) if the training set sufficiently covers the variability of the target domain.

With the initial conceptualization of the colorization process examined through the overfitted data processing, the next step was generalization. Hence, we expanded our experiments to assess the model's generalization performance. Approximately, 15000 training patches and 6000 validation patches were used in the experiment. The model was tested exclusively on unseen TA-PARS images, which were taken from the same label-free tissue sections. Again, the results show strong visual agreement between the TA-PARS virtual staining and the H&E ground truth, where the colorization of different tissue structures resembles what is usually seen in H&E-stained tissue samples. Specifically, of interest in the tissues shown in Fig. 7, are the nuclei. In many malignancies, the morphology and distribution of individual nuclei may be indicative of the progression and severity of the disease. Hence, accurately representing the nuclear structures is paramount for any virtual staining technique. As demonstrated in Fig. 7 there is high acuity in the replication of the nuclei shape and size between the colorized TA-PARS and the true H&E. Assessing the enhances section in Fig. 7(a), the resemblance is even more apparent as the nuclear structures are highly similar between the H&E and the virtual staining.

In addition to the visual agreement, two statistical metrics were used for comparison, the structural similarity index measure (SSIM)[40] and the root mean square error (RMSE). The colorized and ground truth images were converted to the Lab color space prior to SSIM and RMSE computations to represent the images in a perceptually correlated color space. The RMSE was chosen to provide a direct comparison between the images, capturing true differences between the virtual stain and the H&E. The SSIM was selected as a metric since it measures the perceived change in structural information as well as the contrast and luminance changes, rather than the absolute errors, mimicking the human visual system[40]. This metric (SSIM) is of particular interest due to the subjective evaluation nature of histopathology[41]. The statistical metrics were applied on 1000 patches of size 256 x 256 pixels. The average computed SSIM was 0.91 ± 0.02 while an average RMSE of 14.28 ± 1.61 was measured, indicating objective structural and quantitative agreement between the H&E-*like* TA-PARS and true H&E.

Assessing additional, larger skin tissue sections (Fig. 8), further comparison may be drawn between the diagnostic features of the TA-PARS, and the H&E images. These images capture the dermis of human skin tissue containing predominately connective tissue and blood vessels. Directly observing the raw TA-PARS data (Fig. 8-i) the nuclear structures are discernable in the dark blue/purple, while the connective tissues are seen in shades of yellow and orange. Finally, the microvasculature is distinctly visible in red. Certain details such as the vasculature and nuclei, may be easier to identify and assess in the raw TA-PARS input (Fig. 8-i) as compared to the H&E (Fig. 8-iii) or the virtually stained images (Fig. 8-ii). All the diagnostic information required for a pathologist to perform diagnosis and histological analysis may be present in the raw TA-PARS images. However, the TA-PARS manifestations are colorized significantly different than the gold standard H&E images. Current pathologists have extensive training and pattern recognition skills based primarily on H&E imaging. Hence, substantial retraining would be required for pathologists to interpret the different colors in the raw TA-PARS images. This further motivates the proposed AI colorization technique, which may enable clinical acceptance by transforming the TA-PARS images into a clinically accepted format.

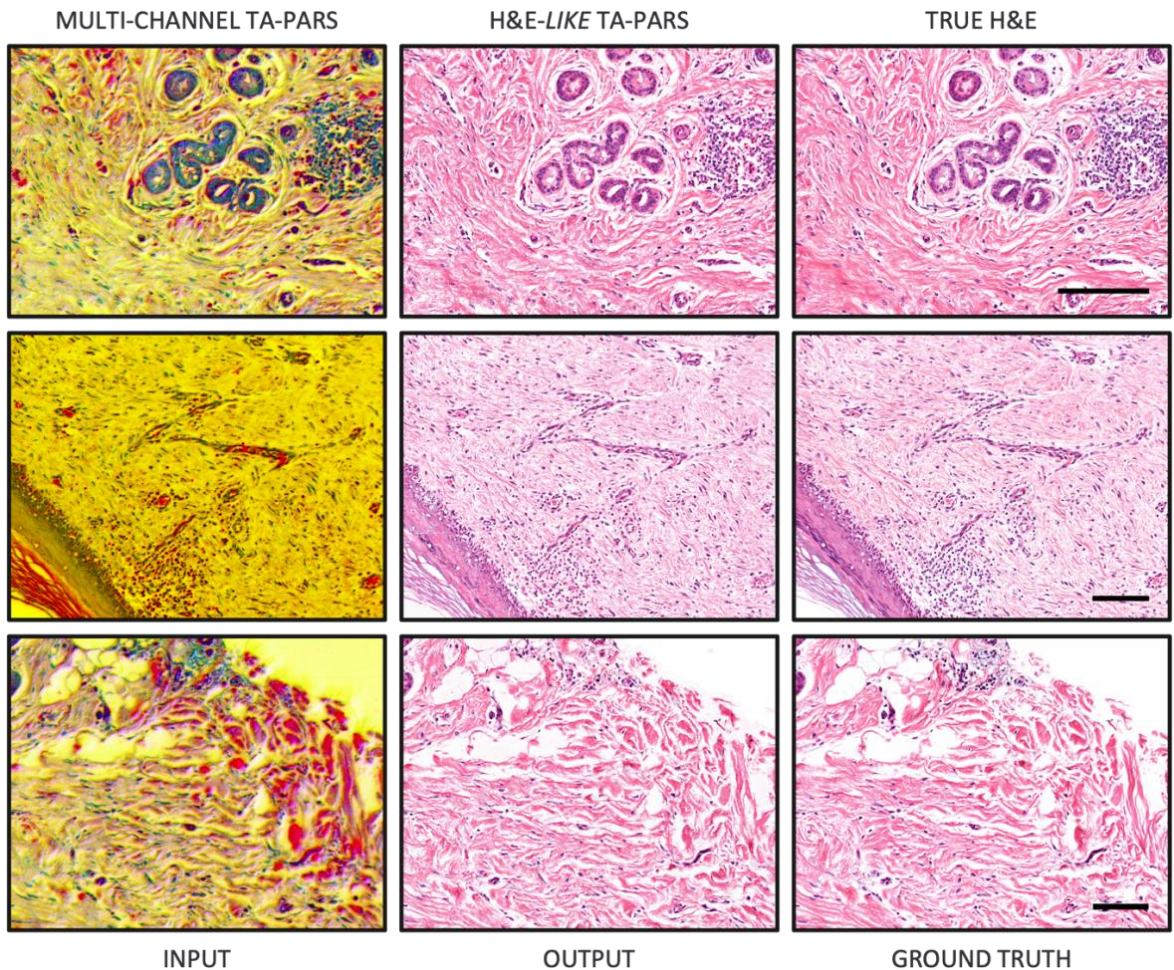

**Figure 6.** Virtual staining of TA-PARS images of human skin tissue slides. Here, the same images were used for training and testing the model, setting an upper bound to the model performance. Scale bar: 100 μm.

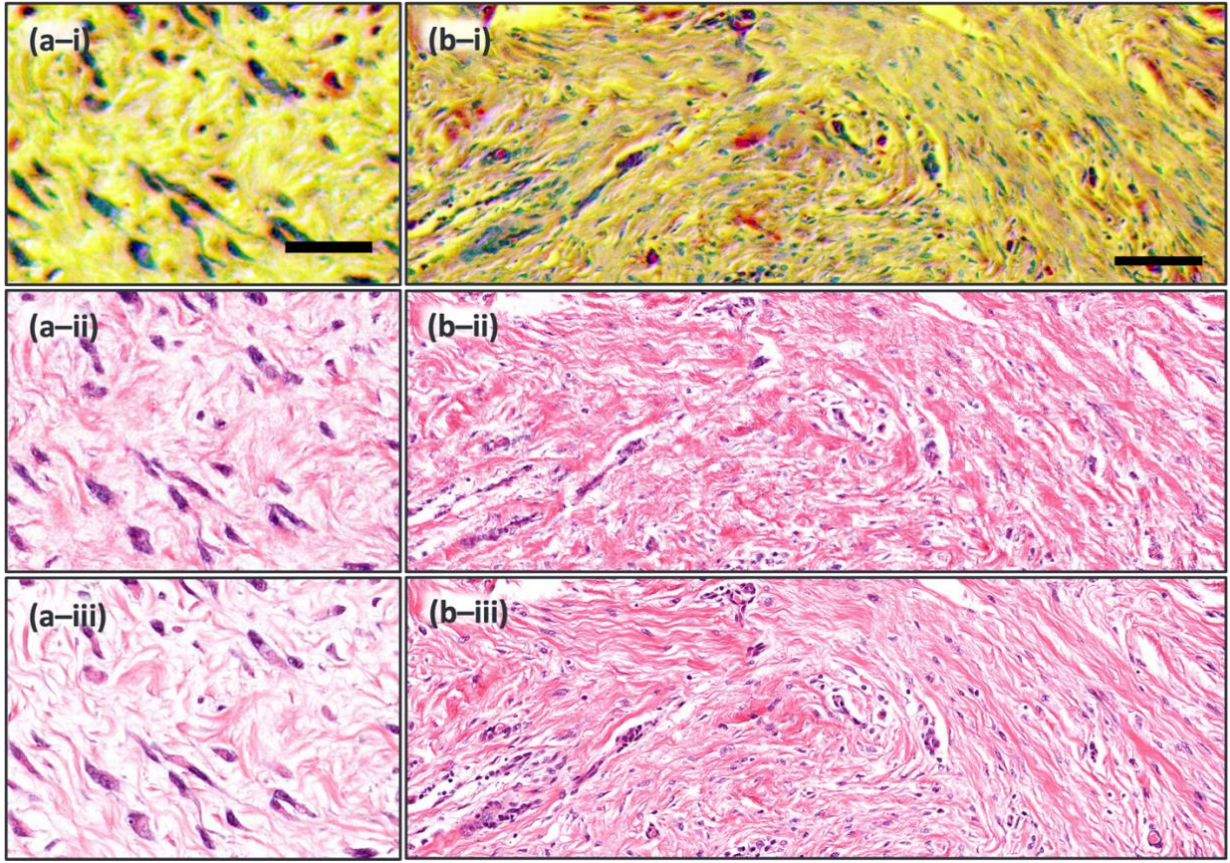

**Figure 7.** Virtual staining results of different parts of human skin tissue using unseen data. (a) Magnified image of connective tissue. Scale bar 25 µm. (b) Image of dense irregular connective tissue. Scale bar 100 µm. (i) Raw TA-PARS. (ii) H&E-like TA-PARS. (iii) True H&E. Image sets (a) and (b) depicts high visual agreement, especially of nuclear structures, between the H&E and the virtual staining.

With TA-PARS virtual H&E, both the collagen rich subcutaneous connective tissues and nuclear structures are represented with high quality and contrast, effectively comparable to the gold standard imaging. Furthermore, four independent dermatopathologists evaluated the images and confirmed that the colorization, contrast, and resolution are sufficient for clinical diagnosis and subjectively equivalent to H&E imaging. This is an encouraging conclusion as these the TA-PARS images were produced directly from unstained tissue specimens. Though the current study was restricted to tissue sections in order to have one-to-one correspondence with H&E from the same tissue, TA-PARS microscopy also provides high resolution images of freshly resected tissues[34], directly capturing hematoxylin and eosin contrasts and enabling comparable fresh-tissue colorization.

In the near future, formalized clinical studies will be conducted aiming to validate and compare the TA-PARS virtual H&E against traditional visualizations. Once this validation is complete proving the accuracy of the colorizations, the proposed system may be employed directly in bulk resected tissue specimens. When moving to bulk tissues, the TA-PARS imaging provides other potential advantages over standard H&E. In addition to the markedly reduced time to diagnosis, this technique, when applied to fresh tissue, removes numerous pre-analytic variables such as the time since devitalization, time from devitalization to fixation, time of fixation, variability of fixation depending on distance from the surface, and staining variability. This could improve the consistency of histology evaluation by removing operator dependent variability in tissue processing.

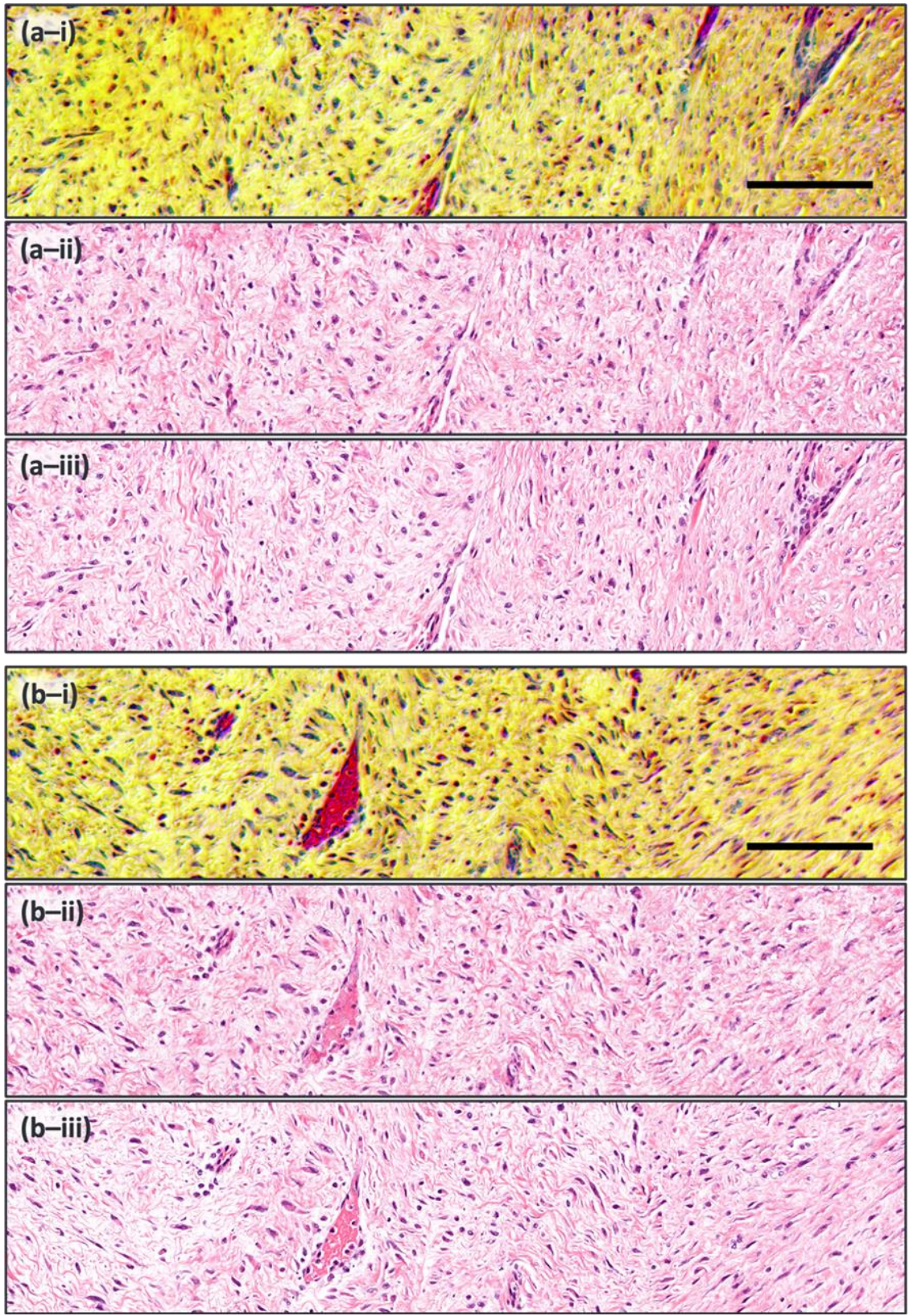

**Figure 8.** Another set of virtual staining results using unseen data of human skin tissue. Image sets (a) and (b) show different parts of the dermis like connective tissue and blood vessels. (i) Raw TA-PARS. (ii) H&E-like TA-PARS. (iii) True H&E. Scale bar 100 μm.

# Conclusion

In summary, a framework was developed for H&E-*like* virtual staining of label-free TA-PARS images. The proposed method leverages the information directly captured from all three TA-PARS channels (radiative absorption, non-radiative absorption, and scattering), providing a robust model which successfully generates H&E-*like* visualizations of tissue samples. Feeding the model with the multi-modal data provides a plethora of data not afforded by any single mode. This may markedly improve the overall colorization quality. Specifically, this method aims to avoid inferring tissue features, to achieve diagnostically accurate colorizations. Applied in thin sections of human skin tissues, the proposed method achieves a high degree of agreement between the virtually staining and gold standard histological staining. Dermatopathologists reviewing the TA-PARS colorized images unanimously confirmed the diagnostic adequacy of these results. Four independent pathologists indicated that the colorization, resolution, and contrast is acceptable as an H&E surrogate. Applied to clinical settings, this may allow for unstained tissues to be imaged, directly producing H&E-*like* visualizations. Such a technique could reduce histological imaging times, while concurrently reducing staining variability by removing operator dependencies within the histochemical staining process. Moving forwards, further exploration will be conducted into different tissue types, and different staining varieties. We envision that this approach will pave the way for fast label-free histological imaging of tissues, which is a key step toward intraoperative microscopic diagnosis and margin assessment.

# Data Availability

Data are available from the authors upon reasonable request.

# Acknowledgements

The authors would like to thank the Cross-Cancer Institute at Edmonton, Alberta for providing human skin tissue samples. The authors also thank Dr. Gilbert Bigras, Dr. Marie Abi Daoud, Dr. Charlene Hunter, and Dr. Karen Naert for their expert pathology review of the TA-PARS images.

The authors thank the following sources for funding used during this project. Natural Sciences and Engineering Research Council of Canada (DGECR-2019-00143, RGPIN2019-06134); Canada Foundation for Innovation (JELF #38000); Mitacs Accelerate (IT13594); University of Waterloo Startup funds; Centre for Bioengineering and Biotechnology (CBB Seed fund); illumiSonics Inc (SRA #083181); New frontiers in research fund – exploration (NFRFE-2019-01012).

# Author Contributions Statement

M.B. implemented the framework, carried out the experiments, prepared the figures, and wrote the main manuscript. B.R.E. collected the TA-PARS images and assisted in writing the manuscript. V.P. assisted in planning the experiments and preparing the results. D.D. and J.R.M. worked on preparing and collecting tissue specimens, provided clinical feedback on the results and performed the clinical consultation in the assessment of the results. P.F. assisted in planning the experiments and provided consultation in manuscript writing. P.H.R. directed and organized the project and the manuscript writing as the principal investigator. All authors reviewed the manuscript.

# Additional Information

## Competing Interests Statement

Authors B.R.E., V.P., D.D., J.R.M., and P.H.R., have financial interests in IllumiSonics which has provided funding to the PhotoMedicine Labs. Authors M.B. and P.F. declare no conflicts of interest.